# Enhancing Oxygen Reduction Reaction on Pt-Based Electrocatalysts through Surface Decoration for Improved OH Reduction Equilibrium and Reduced H$_2$O Adsorption

Yu-Jun Xu,[1] Chiao-An Hsieh,[2] Chen-Yu Zhang,[1] Li-Dan Zhang,[3] Han Tang,[1] Lu-Lu Zhang,[1] Jun Cai,[1] Yan-Xia Chen*,[1], Shuehlin Yau[2] and Zhi-Feng Liu*,[3,4]

[1]Hefei National Research Center for Physical Sciences at Microscale, Department of Chemical Physics, University of Science and Technology of China, Hefei, 230026, China

[2]Department of Chemistry, National Central University, Jhongli 320, Taiwan

[3] Department of Chemistry and Centre for Scientific Modeling and Computation, Chinese University of Hong Kong, Shatin, Hong Kong, China

[4]Ganjiang Innovation Academy, Chinese Academy of Sciences, Ganzhou, Jiangxi, 341119, China

**Abstract:** Electrochemical energy and substance conversion devices involve complex electrode processes, characterized by multiple charge transfer steps, competing pathways, and various intermediates. Such complexity makes it challenging to enhance electrocatalytic activity. The prevailing strategy typically focuses on optimizing the geometric and electronic structures of the electrocatalysts to align the adsorption energies of reaction intermediates with the peak of the activity Volcano curve. In this study, we demonstrate that surface decoration can effectively shape the micro reaction environment for the model system of oxygen reduction reaction (ORR) on Pt electrodes. By applying a partial hydrophobic I* adlayer on the Pt surface, we can shift the equilibrium of OH* reduction and weaken H$_2$O* adsorption, which significantly enhances ORR kinetics. With *in situ* scan tunneling microscopy (STM) and theoretical calculations, our study reveals the formation of isolated Pt$_2$ surface units situated in a hydrophobic valley surrounded by adsorbed iodine atoms. This minimalist Pt$_2$ active unit exhibits significantly greater activity for ORR compared to an extended Pt surface. This strategy could pave the way for developing highly efficient catalysts with potential applications in fuel cell technology and metal air batteries and extension to other electrochemical conversion reactions such as ammonia synthesis and CO$_2$ reduction.

*E-mail: yachen@ustc.edu.cn (Y.X.C.); zfliu@cuhk.edu.hk (Z.F.L.).

# 1. Introduction

Oxygen reduction reaction (ORR) is the primary cathodic reaction in many systems such as the fuel cells, metal air batteries, and electrochemical corrosion of metals. The catalysis of ORR has been a major challenge for decades, as the slow ORR rate imposes a kinetic bottleneck in the advancement of technologies such as fuel cells[1, 2] and metal air batteries.[3, 4] Significant efforts have been devoted to boost ORR kinetics through engineering the electrocatalyst materials and optimizing the ORR conditions.[5, 6] The origins for its slow kinetics and the structure-activity relations of ORR catalysts have also been extensively investigated experimentally.[7-9] Despite over a century of intensive research, the onset overpotential for ORR consistently exceeds 200 mV.[10, 11]

Since 1980s, model studies examining the changes of ORR rates on crystalline metal surfaces have yielded a wealth of kinetic data related to well-characterized surface structures,[10, 12] establishing ORR as a good test ground for validating various theoretical approaches aimed at elucidating electro-catalytic reactions from first principles. The widely used Computational Hydrogen Electrode (CHE) method, which calculates the reaction energy of proton-coupled electrode reactions, was first formulated for the ORR on Pt(111).[13] Based on the thermochemical data obtained through CHE, linear scaling relationships between the binding energies for OOH*, O* and OH*, have been identified, linking the ORR activity to the adsorption energies of these oxygenated species. It produces a volcano curve that accounts for variations in ORR activity across different metals, with Pt, the best monometallic ORR catalyst, at the top of the curve.[6] When applied to more complicated systems, such as bimetallic alloys,[14, 15] Pt shells on metal core substrates,[16-18] and various nano-structured catalysts,[2, 19-22] these volcano curves have provided an effective strategy for enhancing ORR activities by tuning the metal bonding within the electrode, and consequently, the binding energy of oxygenated species. A well-known example is the Pt-skin structure, $Pt_3Ni(111)$, which exhibits 10 times higher ORR activity than Pt(111) surface and a positive shift of half wave potential for ORR by ca. 100 mV, (Figure 1b),[15] making it one of the most effective ORR catalysts to date.

More recent calculations have gone beyond thermochemical values to probe the reaction mechanism. The results highlight the importance of $H_2O$, which is the main component at the ORR cathode/electrolyte interface. As the solvent of aqueous electrolyte and ion transport channels for solid polymer electrolyte membranes, as well as the product of ORR, the concentration of $H_2O$ (ca. 55 M) is more than 4 orders of magnitude higher than that of $O_2$ (ca. 1 mM). In the kinetic region of ORR in acid solution, there is a fast acidic dissociation of adsorbed $H_2O^*$ on Pt(111) through

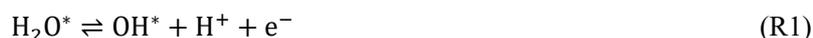
$$H_2O^* \rightleftharpoons OH^* + H^+ + e^- \tag{R1}$$

whose exchange current density is estimated to be around 50 mA cm$^{-2}$.[10, 23] As demonstrated by Density Functional Theory (DFT) calculations, it opens up a hydrolysis reduction channel for O*[24, 25]

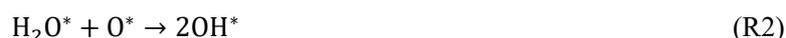
$$H_2O^* + O^* \rightarrow 2OH^* \tag{R2}$$

which is in competition with the protonation of O*, the reverse reaction of the equilibrium

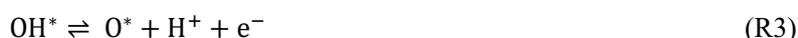
$$OH^* \rightleftharpoons O^* + H^+ + e^- \tag{R3}$$

Recent ab initio molecular dynamics (AIMD) simulations reveal that at elevated electrode potentials, the reduction of O* is dominated by O* hydrolysis (R2) rather than O* protonation (R3).[26] Unfortunately, R2 does not generate electrochemical current, which is responsible for the irreversible oxide region in the cyclic voltammogram shown in Figure 1a. The current associated with 4e$^-$ ORR can only be detected when the applied potential is negative to the $E_{eq}$ for OH* reduction (R1), rather than for O* reduction, as illustrated in Figures 1b,1c and 1d.[10] These mechanistic insights suggest that H$_2$O* is in fact detrimental to ORR, by diverting O* reduction towards the hydrolysis channel, a conclusion supported by recent experiments at the Pt(111)/ionomer interface.[27] This also presents a significant challenge for the enhancement of ORR, as it is difficult to avoid the presence of H$_2$O in an aqueous environment, where it is both the solvent and the product.[28]

In this work, we demonstrate an effective strategy to enhance ORR by shifting the equilibrium of R1 through the decoration of Pt electrode surfaces with I*. Specifically, I* is introduced onto the surface of Pt(hkl), pc-Pt and Pt nanoparticles as a submononlayer. Despite a significant portion of the surface Pt sites being covered by I*, the overall ORR activity is effectively enhanced, with Pt(111)@0.31 ML I* achieving the highest promotion factor, comparable to that of Pt$_3$Ni(111), the best ORR alloy catalyst reported to date.[15] These results challenge the prevalent view that strongly chemisorbed anions, such as I*, diminish the ORR activity.[29, 30] The dependence of the ORR promotion factor on the surface orientation of Pt(hkl) and coverage of I* has been systematically investigated in both acidic and alkaline solutions. The mechanism is elucidated by *in situ* scan tunneling microscopy (STM), Fourier Transform infrared spectroscopy (FTIRS) and DFT calculations, indicating the prospect for simultaneously enhancing ORR and reducing Pt contents.

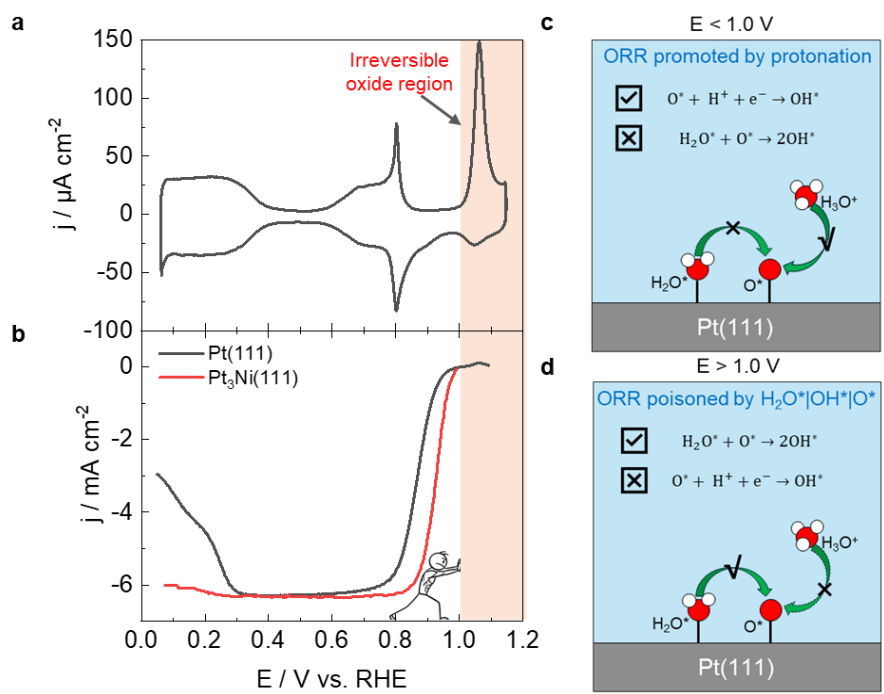

**Figure 1. Oxygen reduction reaction mechanism diagram.** (a) Cyclic voltammogram, (b) representative ORR polarization curves for Pt(111) and Pt$_3$Ni(111) electrode in 0.1 M HClO$_4$, schematic illustration of the mechanisms for the dominating (c) O* protonation channel at E<0.8 V and (d) O* hydrolysis channel at E>0.8 V. j-E curve for ORR at Pt$_3$Ni is adapted from Ref[15] with permission from the publisher.

## 2. Results and Discussion
### 2.1 Electrochemical characterization of Pt(111)@I* electrodes

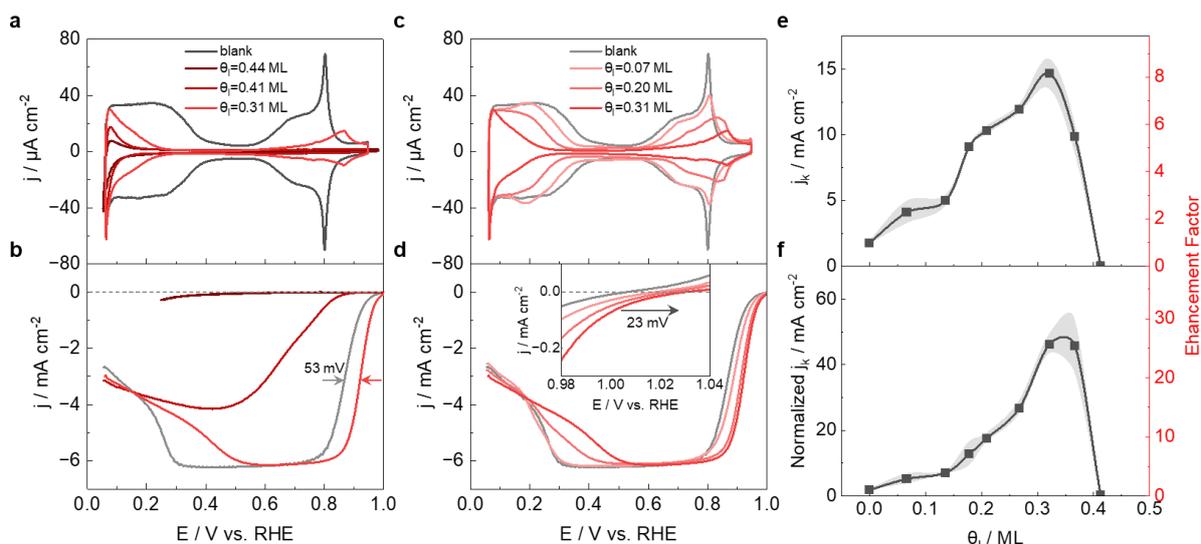

**Figure 2.** Cyclic voltammograms and oxygen reduction reaction activities of Pt@ x ML I*. (**a**, **c**) CVs of Pt(111) electrodes modified by I* with various coverages. (**b**, **d**) Polarization curves for ORR recorded in the positive scan in $O_2$ saturated 0.1 M $HClO_4$. Inset: detailed view of kinetic region of ORR polarization curves and the shift of the onset potential for ORR. The scan rate is 50 mV/s. For the ORR experiments, the electrode rotation rate is 1600 rpm. (**e**, **f**) The $j_k$ of Pt@ x ML I* electrode normalized to the (**e**) geometric area and (**f**) free actives sites as well as the corresponding enhancement factor in reference to that at Pt(111) at 0.9 V. Free active sites are estimated based on the $H_{UPD}$ charge. The shadows indicate error bars in (**e**) and (**f**) represent the standard deviation of three independent measurements.

Shown in Figure 2 are the CVs for Pt(111) partially covered by I* and the corresponding polarization curve at various surface coverages. The iodide adsorption on the Pt electrode surface can be considered as an oxidation process,

$$I^- \rightarrow I^* + e^- \quad \text{(R4)}$$

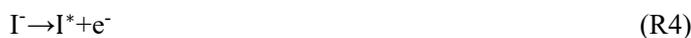

This is supported by DFT calculations that iodine atom is the actual adsorbate on Pt(111), with a small electron transfer (0.08 to 0.13 e/atom) to the surface (see also evidences given in Figure S1 and below).[31, 32] As shown in Figures 2a and 2b, there is almost no ORR activity when Pt(111) surface is fully covered with iodine ($\theta_I$ = 0.44 ML, see also Figure S2), as is well known in the literature.[30, 32] When $\theta_I$ is reduced to 0.41 ML, the partially covered Pt(111) first shows activity towards 2e⁻ ORR, producing $H_2O_2$. Further reduction in $\theta_I$ to 0.31 ML shows enhanced activity towards 4e⁻ ORR process, producing higher current than that on a clean Pt(111) electrode (Figures 2b and 2d). The Pt(111)@I* system thus goes full spectrum, from ORR passivation, to 2e⁻ ORR, to 4e⁻ ORR, with decreasing iodine coverage. This clearly demonstrate that whether I* is a poisoning or a promoter depends on the coverage.

The more interesting change happens in the 4e⁻ ORR region, shown in Figures 2b and 2d. The polarization curve shifts to the rightmost, soon after ORR switches from 2e⁻ to 4e⁻ process around $\theta_I$ = 0.31 ML. Afterwards, the j-E curves for ORR moves gradually to the left with decreasing coverage, towards the curve obtained on a clean Pt(111). All these curves reach a diffusion-limiting plateau for $E$ = 0.3~0.8 V, with the ORR diffusion limiting current ($j_L$) exactly the same as that on clean Pt(111), while both the onset potential (inset in Figure 2d) and the half-wave potential for ORR (Figure 2d) shift positively with increasing $\theta_I$, up to ca. 0.31 ML. Such a trend is quite counterintuitive: the more surface sites of Pt(111) electrode are covered with I*, the better is its ORR activity.

Nonetheless, the corresponding CVs, as shown in Figures 2a and 2c, are roughly similar to that of clean Pt(111). The region with the electrode potential $E$ = 0.05~0.4 V is due to the under-potential deposition of hydrogen (H$_{UPD}$). The region of $E$ = 0.6~1.0 V is due to hydroxyl (OH*) formation through R1. Separating these two is a double layer charging region with $E$ = 0.4~0.6 V. Compared to the CV of clean Pt(111), the onset of H$_{UPD}$ region is pushed to lower $E$ with increasing I* coverage. The decrease in the integral charge density for H$_{UPD}$ in the potential region from 0.4 V to 0.05 V provides a way to estimate $\theta_I$, as is well confirmed that the coverage of H$_{UPD}$ amounts to 2/3 ML at 0.05 V.[33] The calculated values, as shown in Figures 2a and 2c, are in line with those estimated from STM results discussed below. In the OH* region, the current peak for OH* formation is pushed toward higher potentials as $\theta_I$ increases, and the higher I* surface coverage is also indicated by the decrease in the integral charge density for OH* formation.

The ORR kinetic current ($j_k$) free of mass transfer effect can be estimated based on the Kouteckey-Levich Equation:[34]

$$j_k=(j_L \cdot j)/(j_L \cdot j) \quad (1)$$

where j is the measured current density for ORR at a given potential. The Tafel slope can be obtained from the plot of log ($j_k$) vs the electrode potential (see Figure S3). For $\theta_I$ from 0 up to 0.37 ML, the value of Tafel slope is around 50 mV/dec, in agreement with previously reported value for 4e$^-$ ORR on unmodified Pt(111).[35] More exactly, the Tafel slope gradually decreases from 55 mV/dec at $\theta_I$ = 0 to 46 mV/dec at $\theta_I$ = 0.31 ML, as the ORR activity is promoted by partial I* coverage. The ratio of $j_k$ for ORR at 0.9 V versus that on clean Pt(111), is usually taken as the promoting factor for the comparison of ORR activity.[15] As plotted in Figure 2e, the factor is far below 1 at $\theta_I$ = 0.41 ML, when the ORR going through a 2e$^-$ process. It then jumps up at $\theta_I$ = 0.37 ML and soon reaches a maximum at $\theta_I$ = 0.31 ML, with a promotion factor of 8.8, close to the promotion factor ~10 reported on Pt$_3$Ni (111) where the first layer of surface atoms are all composed of Pt (Figure 2e).[15] Furthermore, when normalizing the kinetic ORR current to the available active sites calculated by the reduction in H$_{UPD}$ coverage, the promotion factor for Pt(111)@0.31 ML I* reaches 28.6 (Figure 2f). Thus the normalized ORR activity on Pt(111) partially covered by iodine outperforms that on Pt$_3$Ni (111), the best Pt based ORR catalyst reported so far.[15] When the measurement is conducted in 0.1 M HClO$_4$ solution with 0.1 mM KI, similar to the conditions reported before,[30, 32, 36] we are able to reproduce the results of ORR suppression and I$^-$ oxidation to I$_2$ and IO$_x^-$ above 0.8 V, as Pt(111) is completely covered by I* due to continuous exposure to I$^-$ (Figure S2).

Present findings are in strong contrast with previous reports where it was found that anions in the solution phase, such as halides, sulfate, nitrate, and (bi)phosphate, are often chemisorbed on electrode surfaces, which reduce ORR activities.[12, 37, 38] The general opinion in the literature is that fine tuning the metallic bonds by mixing electropositive elements usually promotes catalysis, while covalent bonding, such as the adsorption of electronegative species, is often deleterious.[30, 39, 40] There are, however, a few intriguing exceptions. Markovic and coworkers demonstrated that the chemisorption of CN$^-$ on Pt(111) enhanced the ORR activity by blocking the adsorption of spectator anions.[41] More surprising were the reports showing the promotion of ORR activity by the moderate coverage of sulfide on Pt/C and Pt black electrodes,[42, 43] even though sulfur is one of the worst poisons for many catalysts. Similar promotion effects were also observed more recently with moderate coverage of Br and Cl on Pt(111).[44] However, the promotion factors in these cases are modest, compared to that achieved on Pt$_3$Ni alloy. The ORR enhancement by CN$^-$ surface patterning on Pt(111) was observed in H$_2$SO$_4$ and H$_3$PO$_4$ solutions, while for the more active ORR in HClO$_4$ solution, there was little improvement over Pt(111).[41] More significantly, there is limited understanding of the underlying mechanisms, even though these results challenge our current conception of catalytic poisons. Without such understandings, it's difficult to formulate strategies for ORR enhancement by surface decorations.

**2.2 Scan tunneling microscopy measurement of Pt(111)@I***

Iodine is also a well-known catalytic poison that passivates Pt at full coverage (Figure 2b), but fortunately, the relevant surface structures have been models for chemisorption and extensively investigated by STM and other surface science techniques.[45-48] To unravel why I* with submonolayer coverage enhances ORR activity, we have carried out *in situ* STM studies to examine the microscopic structure of the iodine adlayer on the Pt(111) electrode in electrochemical environment.

Two samples are prepared and examined separately, for comparison. Sample #1 produces an iodine adlayer with a surface coverage of $\theta_I \approx 0.36$ ML, according to the charge flow in the H$_{UPD}$ process. The surface state of this sample potentiostated at 0.4 V is revealed by STM images shown in Figure 3a. Three patches of ordered arrays (I, II, and III) and scattered disarrays are noted. The symmetry of the structure seen in domain I is defined by the rhombic cell, characterized as $(\sqrt{7} \times \sqrt{7})R19.1°$ with $\theta_I = 0.43$ ML. That of domain II has the same symmetry, but is rotated by 38°. The hexagonal array seen in domain III is characterized as $(\sqrt{3} \times \sqrt{3})R30°$ with $\theta_I = 0.33$ ML. The ball model of the $(\sqrt{7} \times \sqrt{7})R19°$, $(\sqrt{3} \times \sqrt{3})R30°$ are shown in Figures 3f and 3g. Similar adsorbate domain structures has also been confirmed for images taken at 0.7 V or even higher potentials (Figure S5). These two types of iodine structures have been reported before.[45-47]

If these two structures are equally populated, the coverage would be 0.38 ML, slightly higher than 0.36 ML determined by the charge of H$_{UPD}$. The discrepancy is due to the uncounted packing defects in the iodine arrays, imaged as depressions in the STM images. Furthermore, a series of STM images acquired at 0.7 V (Figure S5) reveal that this iodine adlayer was unstable against protracted STM scanning. The ordered iodine arrays could be interconverted and defects migrated on the Pt electrode. Such movements could be too fast to be tracked by the STM, which caused fuzziness in STM imaging.

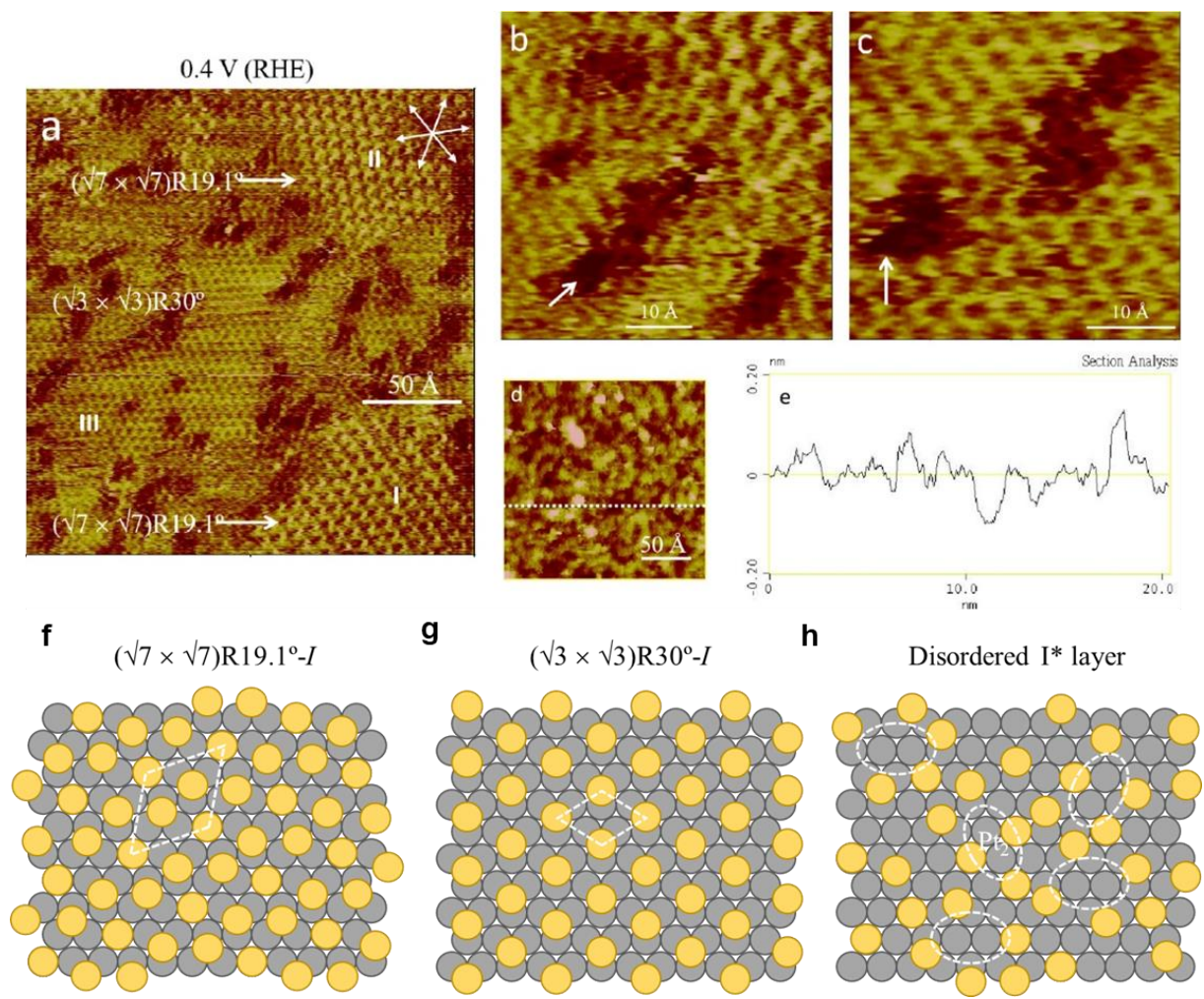

**Figure 3. In situ STM images of Pt@ x ML I*.** In situ STM images showing the structures of iodine adlayer on two Pt(111) samples with high (**a-c**) and low (**d, e**) iodine coverages. The former is mostly covered with ordered ($\sqrt{3} \times \sqrt{3}$)R30º and ($\sqrt{7} \times \sqrt{7}$)R19.1º – I structures, while the latter is disordered. Depressions seen with sample #1 are highlighted in (**b**) and (**c**). Panel (**d**) reveals the general morphology of sample #2. Panel (**e**) shows the corrugation profile along the dotted line marked in (**d**). The potential of Pt is 0.4 V and STM imaging with -200 mV bias voltage and 1 nA feedback current. Panel (**f-h**) is a ball model of the ($\sqrt{7} \times \sqrt{7}$)R19°, ($\sqrt{3} \times \sqrt{3}$)R30° − I and disordered I* layer structure with Pt atoms and I adatoms are colored in grey and yellow, respectively.

Defects in the iodine adlayer on sample #1 are mostly seen at the domain boundaries between the ($\sqrt{7} \times \sqrt{7}$)R19.1º and ($\sqrt{3} \times \sqrt{3}$)R30º or inside the ($\sqrt{3} \times \sqrt{3}$)R30º domains. They are further examined by high-resolution STM scans (Figures 3b and 3c). Only a handful of pits (indicated by arrows) in a 60 × 60 Å scan area are ~1.1 Å lower than the neighboring ($\sqrt{3} \times \sqrt{3}$)R30º array. These are vacancies in the iodine adlayer, which are accessible to oxygen molecules, and subsequently play a vital role in catalyzing ORR. Close examination of Figures 3b and 3c reveals that weak spots ~0.6 Å lower than iodine adatom are present inside these depressions, which could be iodine adatoms in disarrays or ORR intermediates such as O* or OH*.

*In situ* STM scan at 0.4 V in 0.1 M HClO$_4$ solution (open to the air) reveals a completely disordered iodine adlayer, as shown in Figure 3d. Neither switching to a different scan area nor sweeping the potential between 0.1 and 1 V affect the structure of the iodine adlayer. The iodine adlayer has corrugated, rolling-hill morphology with I* aggregating into clusters with poorly defined shapes and dimensions (Figure S6).

Obviously, sample #2 has more vacancy defects in the iodine adlayer than sample #1, which is probably linked to its higher ORR activity, as described above (Figure 2). The cross-section profile shown in Figure 3e reveals the surface roughness of sample #2. Most depressions are ~1.2 Å lower than I* and ~6 Å wide, suggesting that no more than 3 Pt atoms are exposed, since the average Pt−Pt bond distance is ca. 2.8 Å.[49, 50] The possible ball model of $\theta_I$ = 0.33 ML is shown in Figure 3h. The surface roughness of sample #2 becomes more evident, as the corrugation height increases from 1.2 to 2.5 Å upon dosing with $O_2$ (Figure S7). This result is thought to result from the reaction between iodine adatoms and oxygen molecules, producing uncharacterized oxide species on the Pt electrode.

**2.3 Promotion mechanism of Pt(111)@ I* electrode**

As a model system, ORR on Pt(111) has been explored recently by AIMD simulations, with explicit water molecules and real-time treatment of the dynamic effects.[24, 26, 51] These mechanistic understandings are very relevant to the ORR on the Pt(111)@I* system. The key step separating $2e^-$ and $4e^-$ paths is the breaking of the O−O bond. On Pt(111), such a dissociation is facile when $O_2$* is on a side-on configuration (as in Figure S8b), leading to the production of two separate O* atoms and their eventual reduction to $H_2O$ along the $4e^-$ path.[26] It also means that for $4e^-$ ORR, a $Pt_2$ surface unit is required to stabilized these two O* atoms.

The complete passivation of ORR on Pt(111)@I* with $\theta_I$ > 0.44 ML is obviously due to the blocking of all the Pt sites by iodine atoms, as well documented in the literature.[30, 32] The initial reduction of $\theta_I$ from 0.44 ML produces isolated single Pt sites separated from each other, which can now take up $O_2$ in an end-on configuration, rather than the side-on configuration which requires a $Pt_2$ unit. On such a single Pt site, only one O end is bonded to the surface, as shown in Figure S8a, and the dissociation of $O_2$ is prohibited, since a single Pt site cannot accommodate two O* atoms. When it picks up a proton to form OOH*,

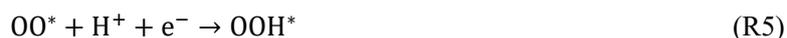

$$OO^* + H^+ + e^- \rightarrow OOH^* \quad (R5)$$

$2e^-$ ORR is initiated. A further step of proton attack would reduce OOH* to $H_2O_2$, and produce the typical $2e^-$ polarization curve, with the current at the diffusion limit being about half of that for the $4e^-$ curve (compare curves in Figure 2b at $\theta_I$ = 0.41 ML and at blank). When $\theta_I$ is reduced below 0.41 ML, $4e^-$ ORR soon becomes the dominant process (at $\theta_I$ = 0.37 ML, Figure S4). The catalytic activity peaks around $\theta_I$ = 0.31 ML. This trend can be attributed to the increasing population of $Pt_2$ sites on Pt(111)@I*, as observed in STM experiments (Figures 3b-3d). With more side-on $O_2$* adsorption structures formed (Figure S8b), the path to O−O dissociation and $4e^-$ ORR is now open.

The I* layer on Pt consists of iodine atoms, and their adsorption on the surface is chemisorption. Tkatchenko *et al* demonstrated by DFT calculations that the iodine-surface bond is covalent, with a chemisorption energy of ~2.6 eV (60 kcal/mol).[31] Interestingly, there is a small charge transfer (< 0.1 e) from iodine to the Pt surface, despite the higher electronegativity of I (2.66) compared to Pt (2.28). This can be explained by the fact that I* is bonded to multiple Pt atom in typical fcc or hcp adsorption configurations, as verified by our own calculations (see Figure S9). Chemisorption I* blocks the active sites on the surface, which is responsible for the catalytic poisoning effect observed at full coverage. However, the covalent bonding transfers a small amount of charge to the surface (Table S1). The d-band center for uncovered surface Pt atom decreases upon I* adsorption by 0.2 eV at $\theta_I$ = 0.31 ML (see Table S2). This change in electronic structure is similar to placing Pt skin on Ni crystals, which decreases the d-band center by 0.1~0.3 eV, depending on the surface index.[15] Thus, while iodine atoms cover some Pt sites by covalent bonding, they do not negatively impact the catalytic activity of the uncovered Pt sites. This is likely also true for the chemisorption of S* or Br*, as their electronegativities (2.58 for S and 2.96 for Br) are close to that of I. It is overly simplistic to claim that covalent bonding on Pt surface inherently poison its catalytic activity.

On the contrary, in the case of Pt(111)@I* systems, iodine-Pt(111) covalent bonding significantly enhances the ORR activity of the uncovered Pt sites nearby. Between $\theta_I$ = 0~0.31 ML, the ORR activity increases significantly as $\theta_I$ is raised (Figure 2e). This observation is counter-intuitive, as increasing coverage reduces Pt sites available for $O_2$ adsorption and its subsequent reduction. The number of available Pt sites for ORR should be the largest at $\theta_I$ = 0 ML. Yet, the ORR activity of a clean Pt(111) is only 1/9 of that at $\theta_I$ = 0.31 ML (Figure 2e) (The intrinsic enhancement of ORR activity after normalization to the number of free active sites is ca. 28.8 times, Figure 2f). Obviously, the presence of I* on Pt(111) promotes ORR, when it's a submonolayer (with $\theta_I$ < 0.41 ML).

There are two aspects for this enhancement effect. The first is the steepening of the ORR polarization curve, as mentioned earlier, with the Tafel slope decreasing from 55 mV/dec at $\theta_I$ = 0 ML to 46 mV/dec at at $\theta_I$ = 0.31 ML. The second aspect is a small shift of the onset potential, by about 23 mV at maximum, as shown in the inset of Figure 2d. Both are related to the overall ORR mechanism, as recently elucidated by AIMD studies, in which the adsorbed $H_2O$* plays a significant role, as mentioned in the introduction. The environment for the adsorption of $H_2O$ is significantly different on clean Pt(111) from that on I@Pt systems.

On clean Pt(111), there are plenty Pt sites for $O_2$*, O* and $H_2O$*. With $H_2O$ being the solvent, it is impossible to avoid $H_2O$* and its poisonous effect through O* hydrolysis. However, a $Pt_2$ site on Pt(111)@I* is surrounded by I* atoms which are almost neutral and do not interact much with $H_2O$ molecules. After the formation of O* atoms by the dissociation O−O* bond, the O* hydrolysis is blocked by the absence of $H_2O$* nearby. It enhances the branching ratio of the O* protonation channel (the reverse of R3) and produces more ORR current. Between $\theta_I$ = 0~0.31 ML, the increase in I* coverage reduces the presence of $H_2O$* and its poisonous effect, which increases the contribution of O* protonation and thus the ORR current, as indicated by the decreases in the Tafel slope, from 55 mV/dec at $\theta_I$ = 0 ML, to 46 mV/dec at $\theta_I$ = 0.31 ML. In this coverage region, the reduction of $H_2O$* poisonous effect is more important for ORR current than the increase of Pt active sites.

The shift in the onset potential is related to R1, the equilibrium between the acidic dissociation of $H_2O$* and the OH* reduction to $H_2O$*. The reverse reaction is required for the completion of ORR. We have calculated the reaction free energy for R1 on clean Pt(111) and on Pt(111)@I* with $\theta_I$ = 0.25 ML, by CHE method,[13] employing an adsorbate solvated by three $H_2O$ molecules, as shown in Figure S10. In terms of the structure, the main difference between Pt(111) and Pt(111)@I* is that two of the $H_2O$ molecules are adsorbed on clean Pt(111), while none of the $H_2O$ molecules are adsorbed on Pt(111)@I*. The calculated ΔG for OH* reduction is -0.55 eV on clean Pt(111), and -0.86 eV on Pt(111)@I*. When the model is expanded to four $H_2O$ molecules, as shown in Figure S11, the free energy is -0.62 eV on clean Pt(111) and -0.68 eV on Pt(111)@I*. In both cases, OH* reduction is more favorable on Pt(111)@I*.

The desorption of H₂O* is also needed to clear Pt sites for O₂ adsorption, although this step itself does not generate current. In this regard, H₂O can again be considered as a catalytic poison, and the weaker its adsorption energy, the better the ORR activity. On a clean Pt(111), H₂O* adsorption is energetically favorable by 0.53 eV, in agreement with values previously reported.[52] This value decreases with increasing number of I* on the surface: 0.52 eV for one I (1/16 = 0.06 ML), 0.45 eV for three I (3/16 = 0.19 ML), 0.25 eV for four I (1/4 = 0.25 ML), and 0.11 eV for five I (5/16 = 0.31 ML). Therefore, both the reduction of OH* to H₂O* and the desorption of H₂O* are facilitated by the weakened H₂O* adsorption on Pt(111) partially covered by I*, which is responsible for the upward shift in the onset potential for ORR, as shown in Figure 2d.

## 2.4 The hydrophobicity of Pt@I*

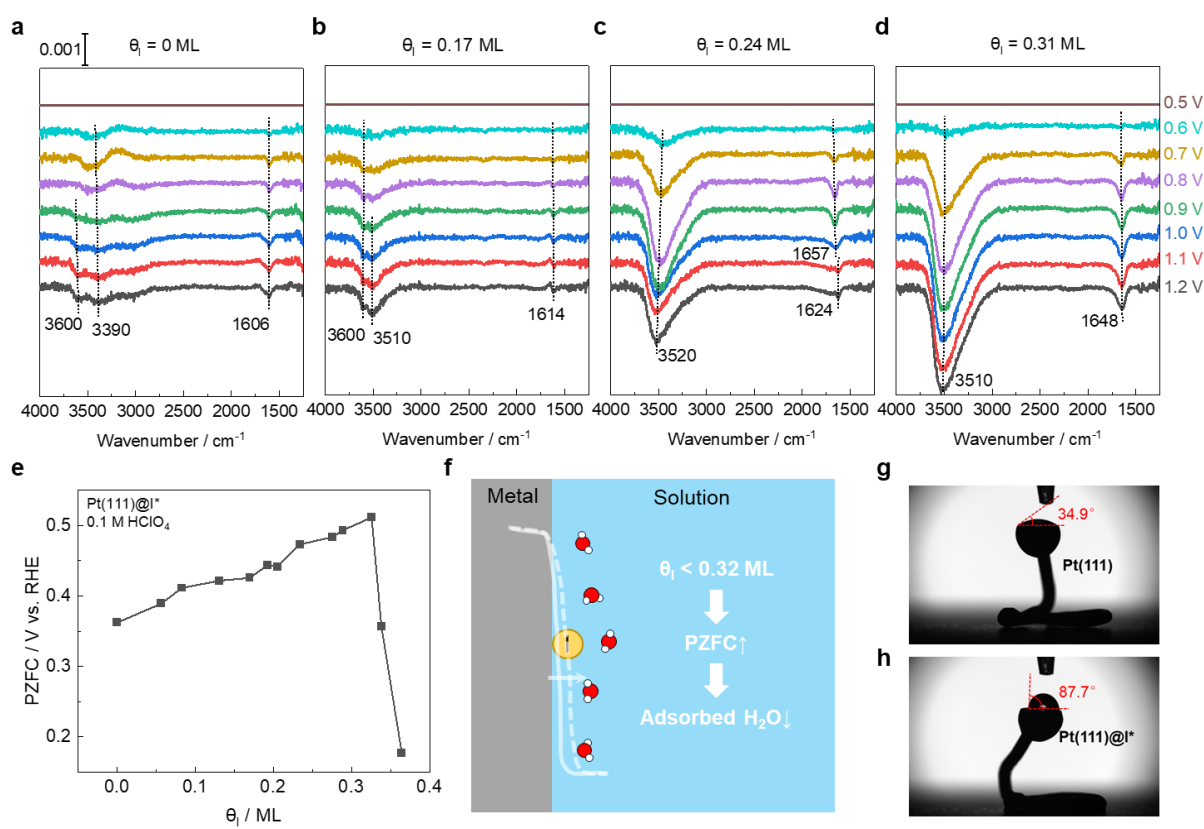

**Figure 4. Evidence for the hydrophobicity of adsorbed I* atoms.** (**a-d**) In situ ATR-SEIRAS spectra were recorded for the Pt working electrode in an Ar-saturated solution at potentials from 1.2 to 0.5 V (vs. RHE) at different I* coverages. The IR background was taken at 0.5 V in the corresponding solution. (**e**) PZFC of the Pt(111)@ x ML I* electrode as a function of I* coverage. (**f**) Schematic illustration of the reduced adsorption of water molecules and the change in electron spillover from the electrode at low I* coverage. (**g, h**) Results of contact angle measurements for Pt(111) and Pt@I*.

In order to verify whether I* mitigates water adsorption at Pt electrode, we have also examined the Pt@I* interface using electrochemical in-situ surface-enhanced infrared absorption spectroscopy (SEIRAS) in the attenuated total reflection (ATR) configuration. Shown in Figures 4a-4d are the spectra at various coverages with $\theta_I$ = 0, 0.17, 0.24 and 0.31 ML respectively, which were estimated from CV (Figure S12). In each case, the spectrum at 0.5 V, shown as a straight line, is taken as the background and subtracted from the original spectrum which contains much contribution from the solvent H₂O molecules. The signals from adsorbed $H_2O^*$ are better revealed in such difference spectra. On polycrystalline Pt thin film electrode (Figure 4a) without I* exposure, small negative peaks are observed around 1606 cm$^{-1}$ for the bending mode and around 3390 cm$^{-1}$ for the O−H stretching mode. The small decrease in these signals at higher electrode potential is due to the acidic dissociation of $H_2O^*$, by R1, to produce OH*, a process well documented in its CV.[53] On Pt electrode with $\theta_I$ = 0.31 ML (Figure 4d), these peaks are significantly more negative, indicating the absence of $H_2O^*$ on iodine covered Pt surface. Comparing Figure 4b at $\theta_I$ = 0.17 ML, Figure 4c at $\theta_I$ = 0.24 ML, and Figure 4d at $\theta_I$ = 0.31 ML, the decrease is more pronounced at higher coverage, consistent with the expectation of less $H_2O^*$ at increased iodine coverage.

The decrease in $H_2O^*$ adsorption is also indicated by the shift in the potential of zero free charge (PZFC). Experimentally, it has been established that $H_2O^*$ adsorption lowers the PZFC for Pt(111), as the electrostatic repulsion between the electrons from oxygen atoms in water and those on Pt(111) impedes the electron spillover across Pt(111)/water interface, and consequently lowers its surface potential (work function).[54, 55] We have measured the PZFC of clean Pt(111) and Pt(111)@ I* with varying $\theta_I$ in $HClO_4$ solution, using $S_2O_8^{2-}$ (Peroxodisulfate, PDS) reduction as the probe reaction (Figure S13), a well-documented procedure for Pt electrodes.[56] A higher and a lower PZFC for the Pt(111)/0.1 M $HClO_4$ interface are identified, and the lower PZFC shift positively with increasing $\theta_I$ up to ca. 0.32 ML (Figure 4e).[57] On clean Pt(111), the lower PZFC is 0.37 V, in agreement with previous reports.[56] Between $\theta_I$ = 0.0 ML, i.e. clean Pt(111), and Pt(111)@I* with $\theta_I$ = 0.32 ML, the PZFC is shifted upward by ca. 150 mV. Considering that adsorbed iodine atoms only transfer a small of amount charge to Pt (0.08 to 0.13 e/atoms),[31] and one adsorbed I* displace several adsorbed water, this increase is likely due to a reduction in $H_2O^*$ adsorption with increasing $\theta_I$ (Figure 4f). Notably, PZFC decreases only when $\theta_I$ exceeds 0.32 ML, as water adsorption reaches a minimum at high I* coverage, making the electronic effect of I* on Pt the dominating factor.[57]

In fact, the hydrophobic nature of Pt(111)@I* can be observed in a simple experiment: the measurement of contact angle. As illustrated in Figures 4g and 4h the contact angle for a water drop is larger on iodine covered Pt(111) than that on a clean Pt(111) surface.

It should be noted that Hoshi and co-workers have systematically investigated the effect of introducing organic hydrophobic groups to Pt crystal electrodes, based on the idea that these groups can affect the water structure near the electrode surface and thus the ORR activity.[58] The promotion factor is moderate for various organic compounds, such as amines,[59] alkanes,[60] and aromatic molecules.[61] However, one tetraalkylammonium cation, THA$^+$ (tetra-n-hexylammonium), stands out with an impressive promotion factor of 8.[62] We were intrigued by this result, since in the same report, it showed that N(alkyl)$_4^+$ cations with shorter alkyl chains, such as methyl, ethyl, or *n*-butyl groups, only produced moderate enhancement. While trying to verify these results, we found that the THA$^+$ used in the report was contaminated by iodide and the promotion effect was actually due to I*, rather than THA$^+$ (See SI for reviewer only). Nonetheless, these organic hydrophobic groups can have moderate enhancement for ORR, likely by a mechanism similar to that discussed above, although with their bulky structures, it's challenging to control the surface coverage and the availability of isolated $Pt_2$ sites for the maximization of promotion effect.

## 2.5 Beyond Pt(111) single crystal electrode and acid media

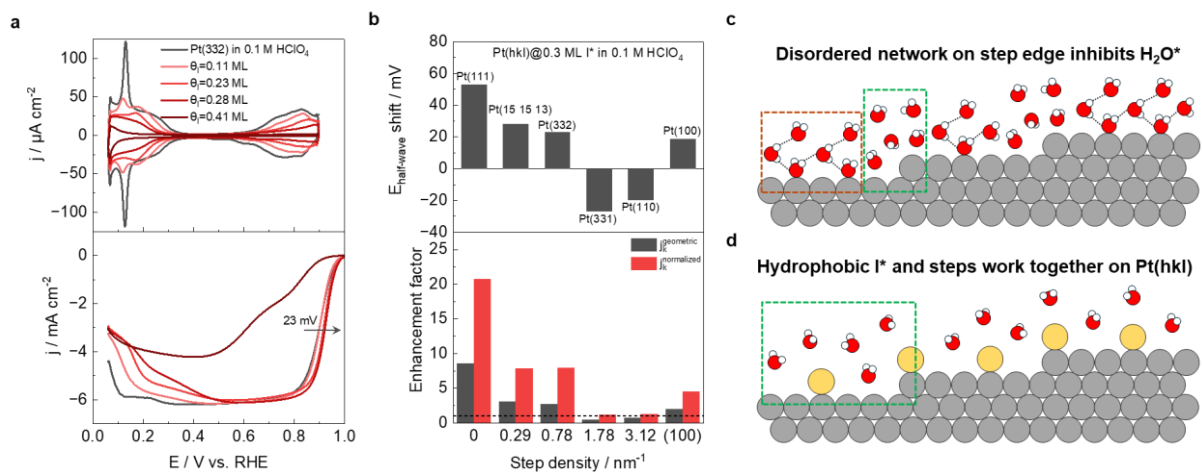

**Figure 5. Activity trends and mechanisms of Pt(hkl)@ x ML I* electrodes.** (**a**) CVs and ORR polarization curves recorded in the positive scan in 0.1 M HClO$_4$ of Pt(hkl)@ x ML I* electrodes. The scan rate is 50 mV/s. For the ORR experiments, the electrode rotation rate is 1600 rpm. (**b**) The shift of half-wave potential and the enhancement factor of j$_k$ at 0.9 V plotted against the step density of Pt(hkl)@ 0.3 ML I* electrodes. (**c**) Schematic illustration of water molecule adsorption on high-index single crystal electrode surfaces with/without I*-modification.

The ORR enhancement by I* submonolayer is also observed on Pt(hkl) surfaces in acid solution. For (111) stepped single crystal electrodes, the ORR activity correlates with the step density, as shown in Figure 5a and 5b. At an I* coverage of 0.30 ML, surfaces with low step density, such as Pt(332) and Pt(15 15 13), exhibit significant enhancement and positive shift in half-wave potential. In contrast, on surfaces with high step intensity, such as Pt(311) and Pt(110), ORR is either less enhanced or even suppressed (see Figure 5b). The presence of (111) terrace is not always required, as enhancement is also observed on Pt(100)@0.30 ML I*, although the promotion factor is smaller than that on Pt(111) (Figure S14). However, no enhancement is observed on polycrystalline Pt electrodes (Figure S15).

For clean Pt(hkl), it has been reported that the ORR activity of clean Pt(hkl) surfaces increases with higher step density, reaching a maximum on Pt(331).[63] This enhancement is attributed to the disruption of the ordered H$_2$O* network by the step edge, which destabilizes H$_2$O* and facilitates the removal of OH* (Figure 5c). However, such a trend is reversed on I* covered Pt(hkl) surfaces, since the presence of I* weakens H$_2$O* adsorption on (111) terraces, as illustrated in Figure 5d, which can promote ORR. I* could also adsorb near the step edge, making its effects on H$_2$O* and ORR an interesting topic for future research.

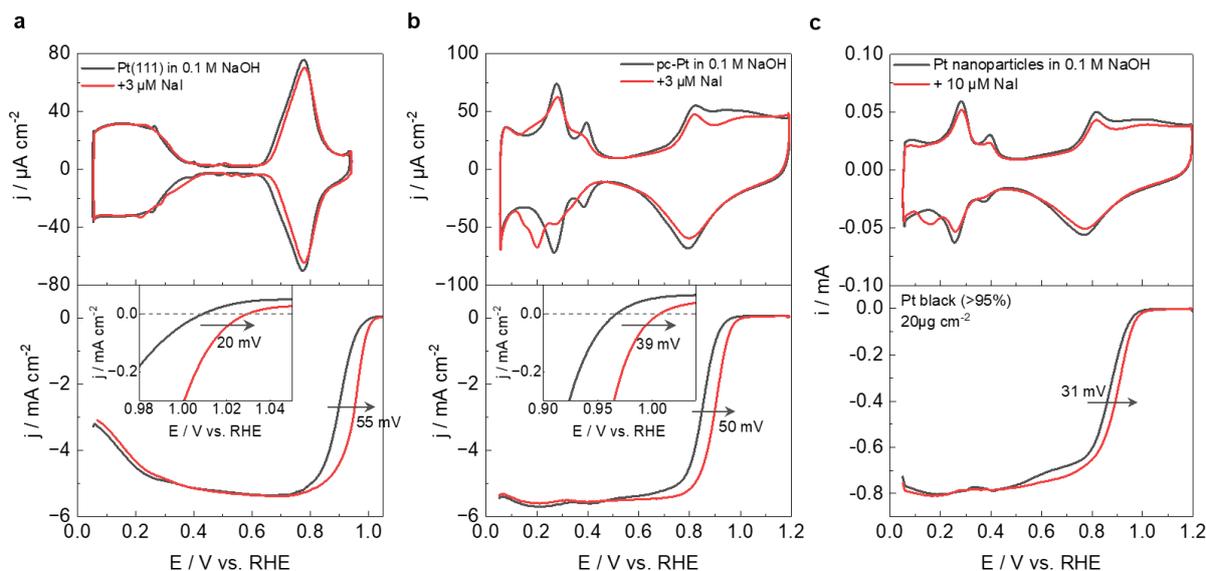

**Figure 6. Activity of Pt electrode in 0.1 M NaOH with/without NaI.** (a) CVs and ORR polarization curves recorded in the positive scan for (a) Pt(111) electrode in 0.1 M NaOH with/without 3 μM NaI, (b) polycrystalline Pt electrode in 0.1 M NaOH with/without 3 μM NaI and (c) Pt nanoparticles electrode in 0.1 M NaOH with/without 10 μM NaI. Inset: detailed view of kinetic region of ORR polarization curves and the shift of the onset potential for ORR. The scan rate is 50 mV/s. For the ORR experiments, the electrode rotation rate is 1600 rpm.

In alkaline solution, similar enhancement of ORR activity is not only observed on I*-modified Pt(111), but also on electrodes composed of polycrystals and nanoparticles Pt, as shown in Figure 6. The half-wave potential for ORR shows a positive shift upon the addition of NaI, by 55, 50, and 31 mV, and the promotion factor is ca. 4.0, 6.2 and 2.2 for $j_k$ at 0.9 V, respectively. The ORR mechanism on Pt(111) in alkaline media has also been studied by AIMD recently, showing a 4e⁻ pathway similar to that in the acid solution.[51] The symmetric peaks at 0.78 $V_{RHE}$ in Fig.6a can be attributed to the OH* adsorption/desorption,

$$OH^* + e^- \rightleftharpoons OH^- \qquad (R6)$$

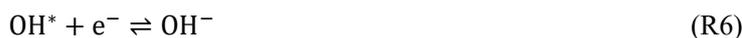

corresponding to (R1) in the acid case. O* reduction can also produce current by

$$O^* + H_2O + e^- \rightleftharpoons OH^* + OH^- \qquad (R7)$$

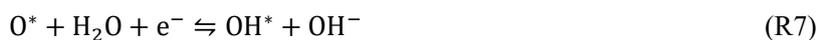

corresponding to (R3) in the acid case. However, above the equilibrium potential of (R6), OH⁻ would be adsorbed on Pt(111) via the reverse of (R6). R7 followed by the adsorption of OH⁻ leads also to an overall reaction of O* hydrolysis (R2), with no net electrochemical current produced. The onset potential for ORR is again determined by OH* reduction (i.e. desorption), rather than by O* reduction, and the presence of $H_2O^*$ is also deleterious for ORR in alkaline media. Furthermore, OH* desorption is not achieved by the breaking of Pt−OH* bond. Rather it is facilitated by the formation of Pt−OH₂*, which leaves a solvated OH⁻ in the outer sphere through proton migration.[51] The desorption of $H_2O^*$ is again the final step of ORR. Therefore, it's not surprising that the hydrophobic I* improves the ORR performance by suppressing R2 and promoting R7 in alkaline solution, shifting up the ORR onset potential by 20 and 39 mV on Pt(111) and polycrystalline Pt (inset in Figures 6a and 6b).

While in acid solution the enhancement effect of I* is favored on (111) terrace, in alkaline solution, it is observed on all morphological forms. This lack of facet selectivity may be attributed to the preferential adsorption of hydroxyl anions on edge sites, which leaves I* on the flat terraces and weakens H$_2$O* adsorption in those areas as well.[64] These results demonstrate that surface decoration can be an effective ORR enhancement strategy on Pt(hkl), polycrystal-Pt and nanoparticle Pt electrodes, especially in alkaline solution.

## 3. Conclusions

We have shown that Pt electrodes partially covered by iodine can enhance ORR activity, with a promotion factor close to the performance of the best alloy ORR electrode, Pt$_3$Ni(111). With I* coverage systematically varied, the link between the decrease in H$_2$O* adsorption and the increase in ORR activity is established by experimental measurements on Pt(111), especially by in-situ STM and in-situ vibrational spectra and supported by DFT calculations. These results demonstrate one possible answer to the challenge of H$_2$O* poisoning on electrode surface during ORR, predicted by recent mechanistic studies. By embedding active Pt$_2$ sites in the valley of I@Pt and protecting these sites from solvent H$_2$O molecules, it is possible to achieve a more than eight-fold increase in ORR activity, despite the ~70% reduction of active Pt sites.

Such a catalyst model for ORR runs counter to two commonly held assumptions. First, it is often assumed that the formation of covalent bonds between non-metal atoms with Pt surface is deleterious for its catalytic activity. I* is indeed a well-known catalytic poison for ORR on Pt, as the chemisorption of iodine atom on Pt can easily take up all the Pt sites. However, our study shows that the bonding interaction between I* and Pt surface in itself does not poison the nearby Pt sites. In fact, it promotes ORR, because Pt has a relatively large electronegativity among metals and I* actually donates a small amount of charge to Pt surface. Furthermore, the presence of I* on partially cover Pt(111) lessens H$_2$O* adsorption on uncovered Pt sites and enhances ORR by reducing the poisonous effect of H$_2$O*. This is likely also the mechanism for the ORR enhancement previously observed on Pt electrodes partially covered by S* and Br*, indicating the potential of using even non-metal atoms either in surface decoration or as substrate to tune the OH* reduction energy and the H$_2$O* adsorption energy.

Second, it is also often assumed that increasing surface area would enhance catalytic activity. But in the case of ORR, extended Pt surface makes it hard to avoid H$_2$O* adsorption. Four electron ORR needs Pt$_2$ as its basic catalytic unit, so that both O* atoms in O$_2$* or OOH* interact with Pt atoms and O−O bond breaking is facilitated. Nearby Pt sites would attract H$_2$O* adsorption, which reduces ORR current. Instead of extended Pt surface, an improved strategy is to create an ensemble of active Pt$_2$ units, isolated from each other and situated on non-Pt substrate surrounded by hydrophobic adsorbates, as shown in Figure S8d. Such a catalyst can enhance the ORR activity by mitigating water poisoning, while lowering the platinum content to a minimum, both being much desirable goals in the advancement of fuel cell technology.

**Acknowledgments**

This work was supported by the National Natural Science Foundation of China (no. 22372154, 22172151), by the Research Grants Council of Hong Kong SAR Government (GRF Grant 14303114), by the Ministry of Science and Technology of the People's Republic of China (G2022200006L), and by the National Science and Technology Council of ROC (NSTC 112-2113-M-008-005). We are grateful for the generous allocation of computer time on the HPC clusters at the Center for Scientific Modeling and Computation, CUHK.

Supporting Materials for

Enhancing Oxygen Reduction Reaction on Pt-Based Electrocatalysts

through Surface Decoration for

Improved OH Reduction Equilibrium and Reduced H$_2$O Adsorption

Yu-Jun Xu,[1] Chiao-An Hsieh,[2] Chen-Yu Zhang[1], Li-Dan Zhang,[3] Han Tang,[1] Lu-Lu Zhang,[1] Jun Cai,[1] Yan-Xia Chen*,[1], Shuehlin Yau[2] and Zhi-Feng Liu*,[3,4]

[1]Hefei National Research Center for Physical Sciences at Microscale, Department of Chemical Physics, University of Science and Technology of China, Hefei, 230026, China

[2]Department of Chemistry, National Central University, Jhongli 320, Taiwan

[3] Department of Chemistry and Centre for Scientific Modeling and Computation, Chinese University of Hong Kong, Shatin, Hong Kong, China

[4]Ganjiang Innovation Academy, Chinese Academy of Sciences,
Ganzhou, Jiangxi, 341119, China

*E-mail: yachen@ustc.edu.cn (Y.X.C.); zfliu@cuhk.edu.hk (Z.F.L.).

**Supporting results**

**1. Facts that support I* adatoms at Pt(111)@x ML I* are almost electroneutral**

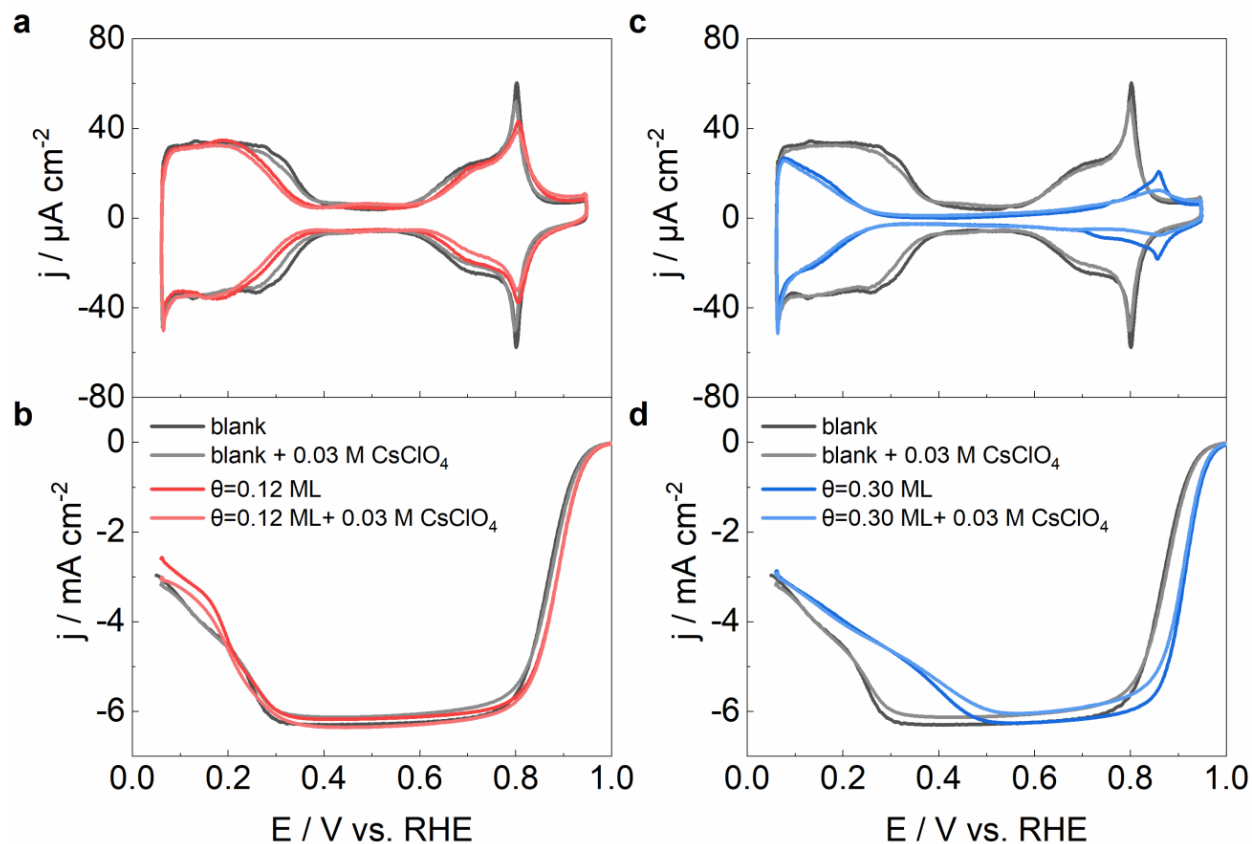

**Figure S1.** (**a**, **c**) CVs and (**b**, **d**) positive-going j-E curves for ORR at Pt(111)@x ML I*/0.1 M HClO$_4$ interfaces with and without addition of 0.03 M CsClO$_4$. For comparison, the base CV and the j-E curve for ORR at bare Pt(111) are also included in (**a**, **c**) and (**b**, **d**). The scan rate is 50 mV/s. For the ORR experiments, the rotation rate is 1600 rpm.

The base CVs of Pt@ x ML I* as well as the j-E curves for ORR with and without addition of CsClO$_4$ is nearly the same, this further supports that I* is nearly neutral, electrostatic interaction among I* and Cs$^+$ is negligibly small. DFT calculation on the charge transfer for I* at the fcc site of Pt, determined from the Hirshfeld partitioning method, also suggests that the iodine atom remains almost neutral upon adsorption.[1,2,3] This is in good agreement with previous report that I* at Pt(111) are largely neutral atoms, as evident by the negligible amount of ions is found in the thin liquid film after the electrode is immersed out from the electrolyte.[4] This is also supported by the fact that when Cs$^+$ is added into the electrolyte, ORR activity does not show obvious change (Figure S1).

It is well confirmed by the fact that there is significant current flow through the external circuit when such electrode is pulled out from the electrolyte with 0.1 mM KI.[4] The current comes from the discharging of the charges initially stored in the electrochemical double layer (EDL) when the EDL breaks due to the pulling action. In contrast, for other species with anions as their adsorbing state, due to strong electrostatic interactions, liquid film with the whole EDL will be dragged out upon immersion out the solutions, under such circumstances, since there is no breakage of EDL no charge flows through the external circuit.

## 2. The poisoning effect of halogen on ORR kinetics of Pt(111) electrode

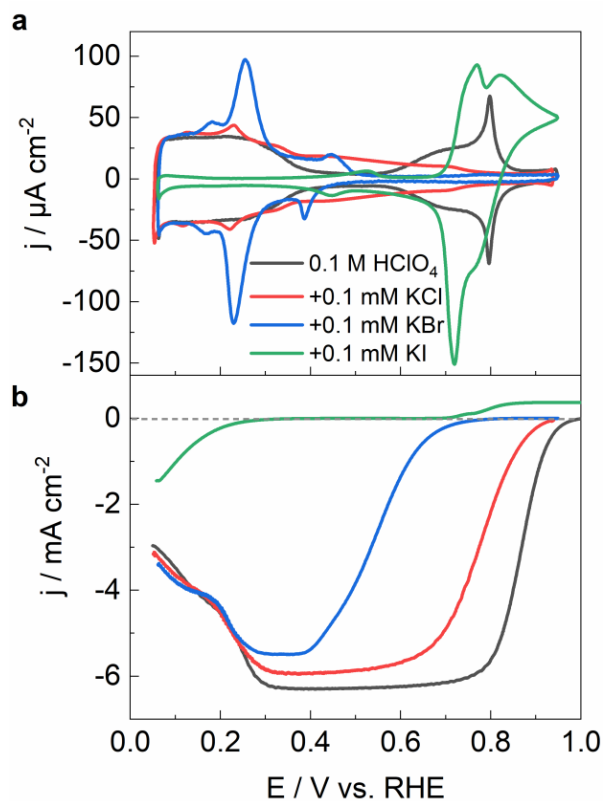

**Figure S2.** (**a**) CVs of Pt(111) electrodes and (**b**) positive-going j-E curves for ORR in 0.1 M HClO$_4$ + 0.1 mM KX (X= Cl, Br, I),. For comparison, the base CV and the j-E curve for ORR at bare Pt(111) are also included in (**a**) and (**b**). The scan rate is 50 mV/s. For the ORR experiments, the rotation rate is 1600 rpm.

Figure S2 shows the CVs and ORR polarization curves in 0.1 M HClO$_4$ with addition of 0.1mM KX (X=Cl, Br, I) respectively. The j-E curve shifts significantly to the negative potentials even though only 0.1 mM KX is added into the solution. With addition of KX with the same concentration, the extent for the inhibition of ORR at Pt(111) decreases in the order of KI > KBr > KCl, this is in good agreement with previous reports.[3, 5, 6]

In solution with 0.1 M HClO$_4$ + 0.1 mM KI, Pt(111) surface is fully saturated with I* in the potential regime from 0.3 V to 1.0 V. In O$_2$ free solution, the anodic current at E>0.7 V is from the oxidation of I$^-$ to I$_2$, subsequently I$_2$ combines I$^-$ to form I$_3^-$ (3I$^-$ ↔ I$_3^-$+2e$^-$). The cathodic current in the same potential regime is due to reduction of I$_3^-$ to I$^-$ (I$_3^-$+2e$^-$ ↔ 3I$^-$), as in good agreement with previous reports.[7] .

The pair of small redox peak at ca. 0.5 V is also quite similar to previous observation, is probably related to the phase transition of I* adlayer.[8] In O$_2$ containing solutions the redox between I$_3^-$ and I$^-$ are superimposed with ORR, since ORR current is tenth of times higher, its contribution is masked.

Since in solution with 0.1 M $HClO_4$ + 0.1 mM KI, Pt(111) surface is fully saturated with I* in the potential regime from 0.3 V to 1.0 V, no ORR current is observed at E>0.3 V when this solutions is saturated with $O_2$. At E<0.3 V, cathodic current due to partial reduction of $O_2$ to $H_2O_2$ can be observed. This is probably due to desorption of some I*, which leads to free active sites for $O_2$ adsorption and reduction.

In contrast for the case with 0.1 M $HClO_4$ + 0.1 M KBr, desorption of Br* is seen at potentials below 0.5 V. For the case with 0.1 M $HClO_4$ + 0.1 M KCl, desorption of Cl* is seen at potentials below 0.8 V. Accordingly, comparing to the case with Pt(111)@saturated I* adlayer, ORR current at the same potential is less poisoned by Cl* and Br*.

## 3. The variation in the Tafel slope of ORR at Pt(111)@I* with changes in I* coverage

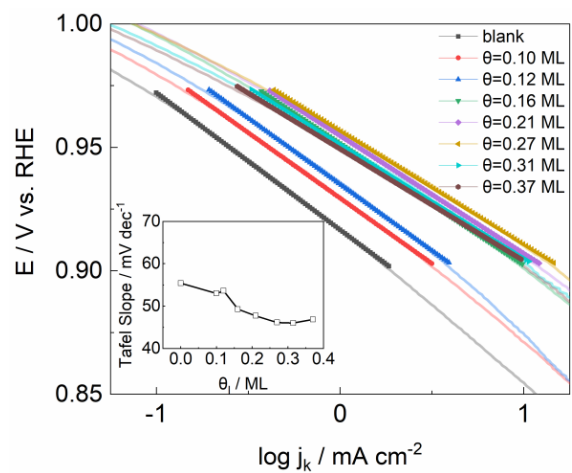

**Figure S3.** Tafel plots of polarization curves for ORR at Pt(111)@x ML I* / 0.1 M $HClO_4$ interfaces recorded in the positive scan. The point lines are fits of the curves from which the indicated Tafel slopes are obtained. Inset: the relationship between ORR Tafel slopes and I* coverages.

The Tafel slope for ORR has a value around 55 mV/dec at blank Pt(111) (Figure S3) in acidic media, in agreement with previously reported value for ORR on clean Pt(111). The Tafel slope decreases with increasing I* coverage, reaching 46 mV/dec at 0.31 ML. The change in Tafel slope indicates that the ORR kinetics increase with rising coverage, suggesting a stronger preference for the protonation channel over the hydration channel as coverage increases.

**5. EC-STM reveals I* adlayer at Pt(111) has defined domain structures as well as defects between domain boundaries, the latter probably acts at ORR active sites**

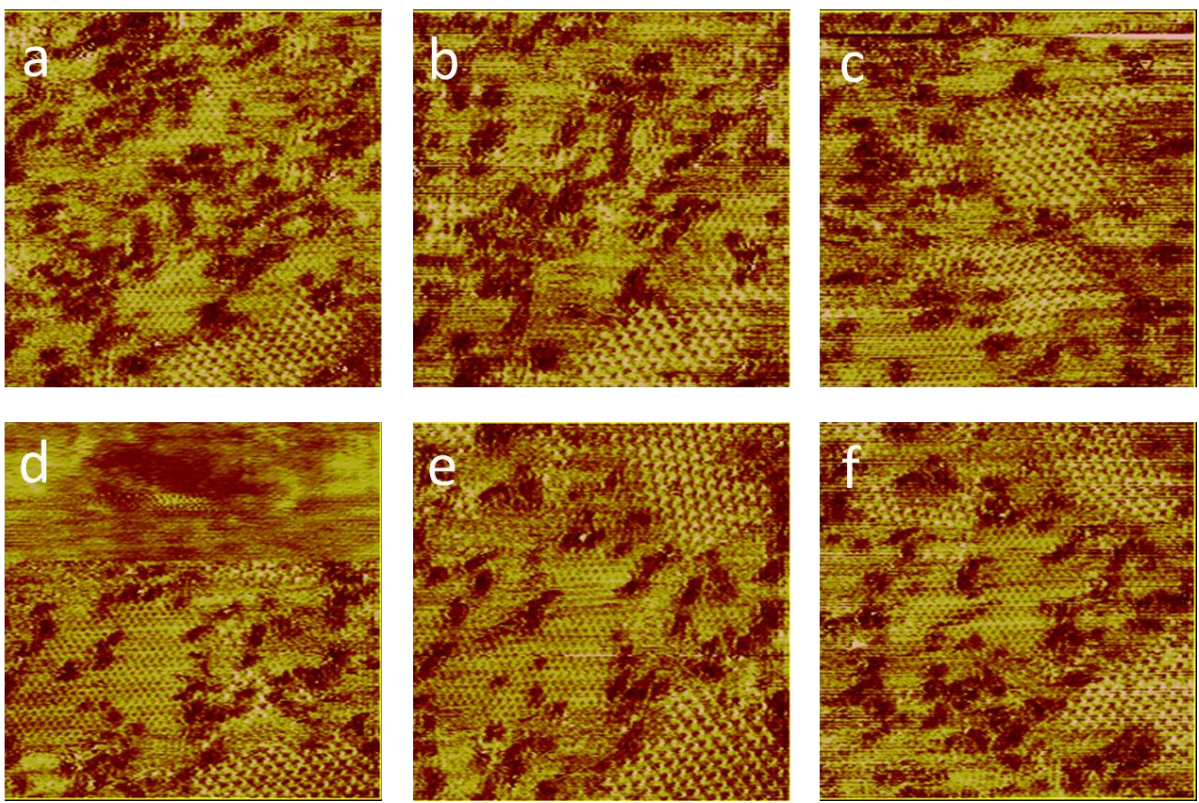

**Figure S5.** A series of STM images showing the constant restructuring of the iodine adlayer on the Pt(111) electrode at 0.7 V in 0.1 M HClO$_4$. The STM are collected consecutively at the same area on the Pt(111) sample in a time interval of 25 s. Scanning with a bias potential between the tip and substrate is -300 mV and a tunneling current of 1 nA. All images are 200 × 200 Å. The abrupt change seen at the upper half of panel (**d**) could be caused by migration of iodine adatoms on the Pt electrode.

The STM images in Figure S5 reveal that there is a brighter ($\sqrt{7} \times \sqrt{7}$)R19.1° and a dimmer ($\sqrt{3} \times \sqrt{3}$)R30° iodine adlattices in these images, they are also domains where I* adlayer do not form well-ordered structures. From the series of STM images, it can also be seen that the I* adlayer structure changes with time dynamically. This agrees well with the fact that the interaction of I* with Pt is not strong.[2, 3]

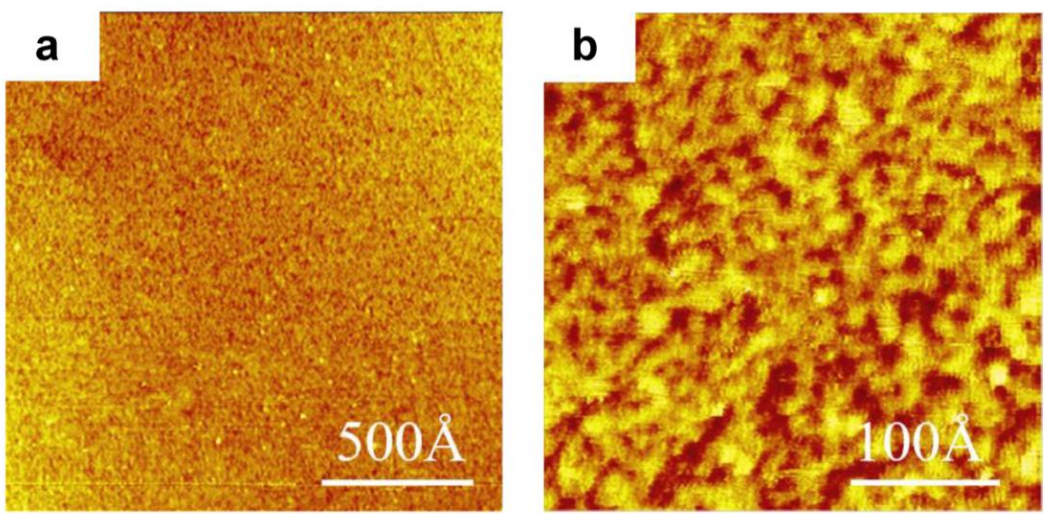

**Figure S6.** A large-scale of STM images showing the disordered covered iodine adlayer on the Pt(111) electrode at 0.62 V in 0.1 M HClO$_4$.

Shown in Figures S6a and 6b are STM images obtained at 0.62 V in 0.1 M HClO$_4$, which reveals that the Pt surface was universally and uniformly covered with an iodine adlayer. A finer resolution STM image shown in Figure S6b reveals that the iodine adatoms formed clusters ranging from 1 to 5 nm. It is difficult to obtain atomic resolution STM image of this iodine coated Pt(111) sample.

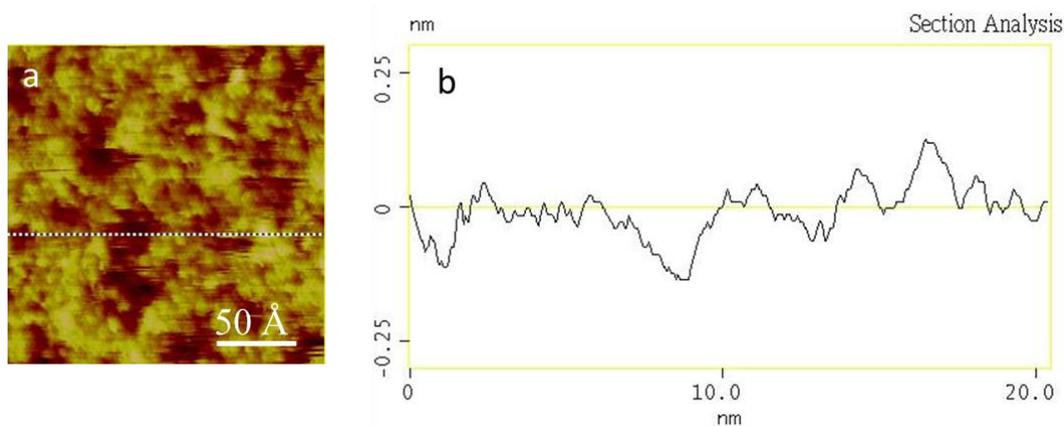

**Figure S7.** In situ STM image (**a**) and cross section profile (**b**). The Pt electrode was set at 0.7 V in O$_2$-saturated 0.1 M HClO$_4$ and tunneling condition is -200 mV and 1 nA. The Pt electrode surface became more corrugated with a maximal height difference of 4 Å, implying that iodine adatoms reacted with oxygen to produce uncharacterized oxide species.

The surface morphology of sample #2 exhibits significantly increased roughness upon O$_2$ exposure, as evidenced by the rise in corrugation height from 1.2 Å to 2.5 Å (Figure S7). This pronounced change suggests a chemical reaction between adsorbed iodine atoms and oxygen molecules, leading to the formation of unidentified oxide species on the Pt electrode surface.

## 6. The structure of O₂ and H₂O adsorption on Pt surface, the charge density difference and reaction free energy were calculated by DFT.

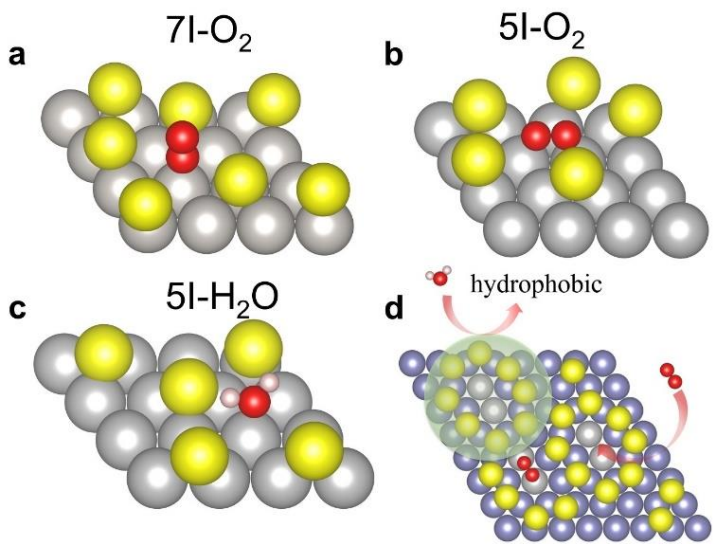

**Figure S8. O₂ and H₂O adsorption structures on Pt(111)@I*, obtained by DFT calculations.** (**a**) end-on O₂* with 7 I* on Pt(111) and $\theta_I = 7/16 = 0.44$ ML; (**b**) side-on O₂*, with 5 I* on Pt(111) and $\theta_I = 5/16 = 0.31$ ML; (**c**) H₂O*, with 5 I* on Pt(111); (**d**) model catalyst for ORR, with isolated Pt₂ units for O₂ adsorption, protected from H₂O adsorption by hydrophobic structure motifs.

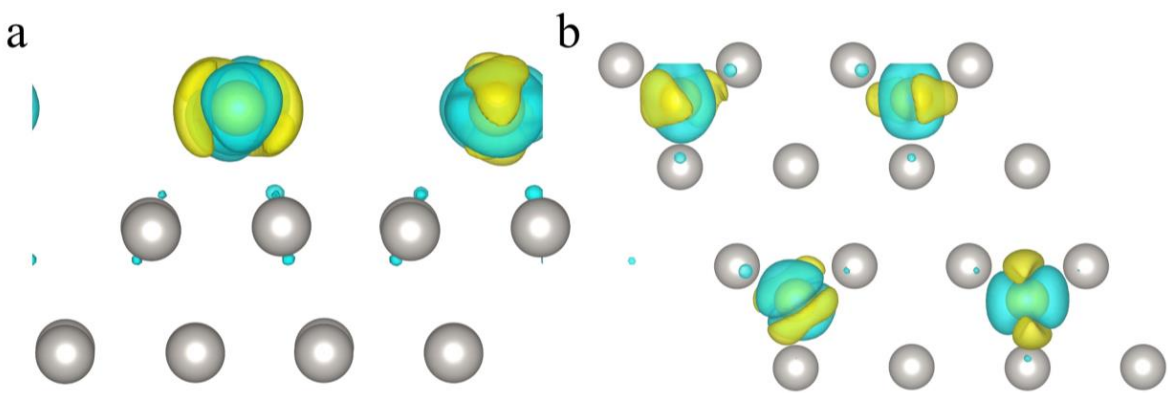

**Figure S9.** The charge density difference of the Pt(111) surface modified by I. (**a**) Top view. (**b**) Side view. The yellow and silver spheres represent the I atom and the Pt atom respectively. The yellow and cyan areas represent electron accumulation and consumption respectively.

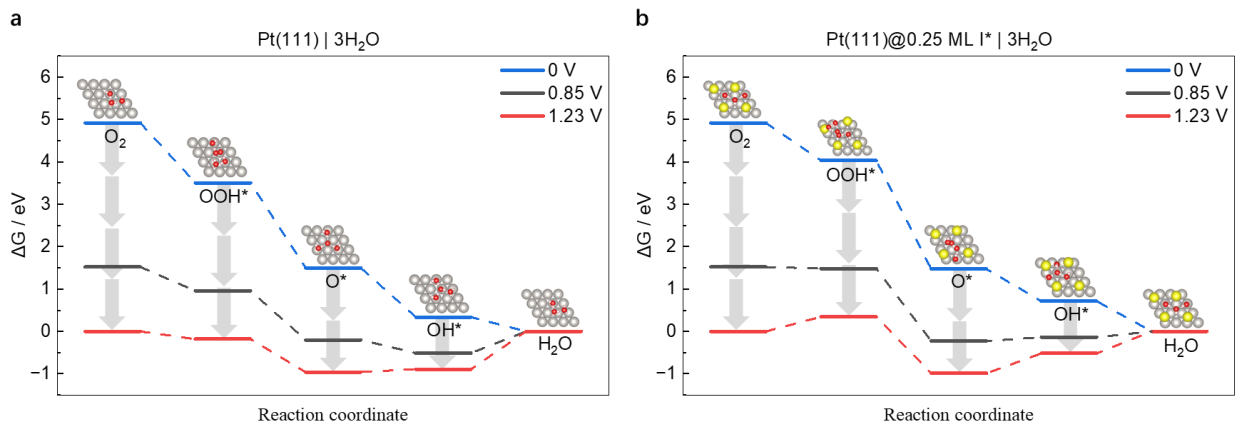

**Figure S10.** Free-energy diagram for oxygen reduction at three different potentials and at (a) Pt(111) | 3H$_2$O interface and (b) Pt(111)@0.25 ML I* | 3H$_2$O interface.

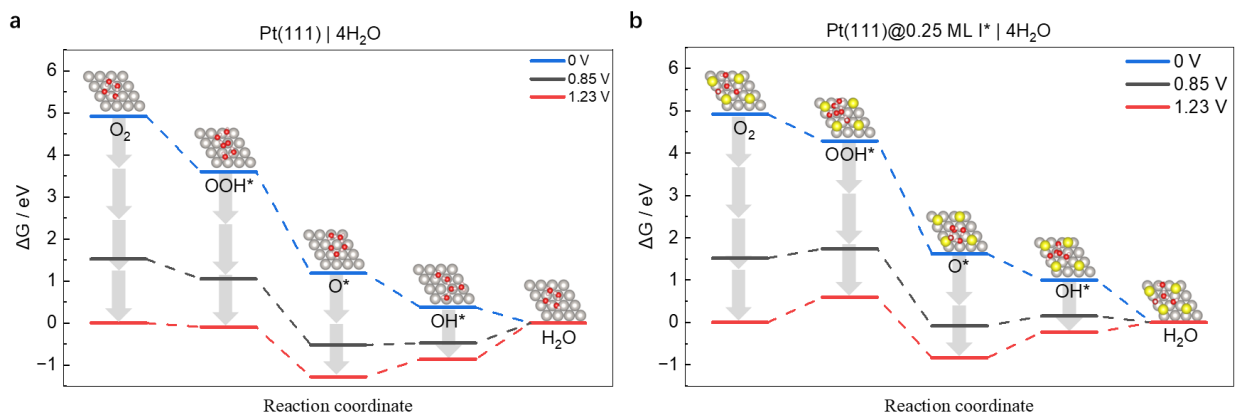

**Figure S11.** Free-energy diagram for oxygen reduction at three different potentials and at (a) Pt(111) | 4H$_2$O interface and (b) Pt(111)@0.25 ML I* | 4H$_2$O interface.

We calculated the reaction free energy for ORR on both clean Pt(111) and Pt(111)@I* with $\theta_I$ = 0.25 ML using the Computational Hydrogen Electrode method,[9] without applying any additional potential correction. In the adsorption models used, the adsorbed intermediates (OOH*, O*, OH* or H$_2$O*) are solvated by three or four explicit H$_2$O molecules (Figure S10 and Figure S11). Structurally, the key distinction between Pt(111) and Pt(111)@I* lies in the adsorption behavior of H$_2$O molecules: two H$_2$O molecules are adsorbed on the pristine Pt(111) surface, whereas no adsorption occurs on the I-modified Pt(111) surface. This microsolvated model captures key features of the hydrogen-bonding network and the first solvation shell at the interface, which have been shown to play an important role in tuning the reaction energetics. Therefore, the free energy changes obtained from this model are meaningful as references for comparing the relative stability of different reaction intermediates, and they provide a rational explanation for the influence of iodine adsorption on the Pt surface in modulating the ORR.

**7. The results of electrochemical in-situ ATR-SEIRAS for the Pt@I* electrode demonstrate the inhibition of I* on the $H_2O$*.**

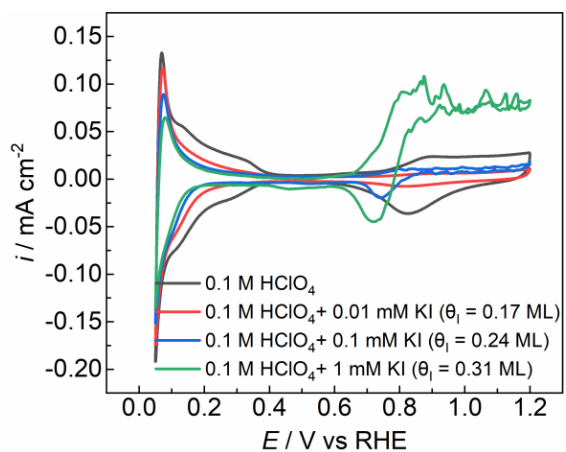

**Figure S12.** CVs of Pt thin film electrodes with different I* coverages in Ar-saturated 0.1 M $HClO_4$ for ATR-SEIRAS test.

Figure S12 displays the CVs of Pt thin film electrodes with different I* coverages in Ar-saturated 0.1 M $HClO_4$ by adding different concentration of KI. Similarly, the $H_{UPD}$ and OH* regions of the Pt electrode decrease with increasing coverage. When 1 mM KI is added, an oxidation peak for $I^-$ can be observed in the OH* region.

## 8. Evidence of positive shift in the potential of zero free charge of Pt(111)@I*/0.1 M HClO$_4$ interface confirms that Pt(111)@I* surface is hydrophobic

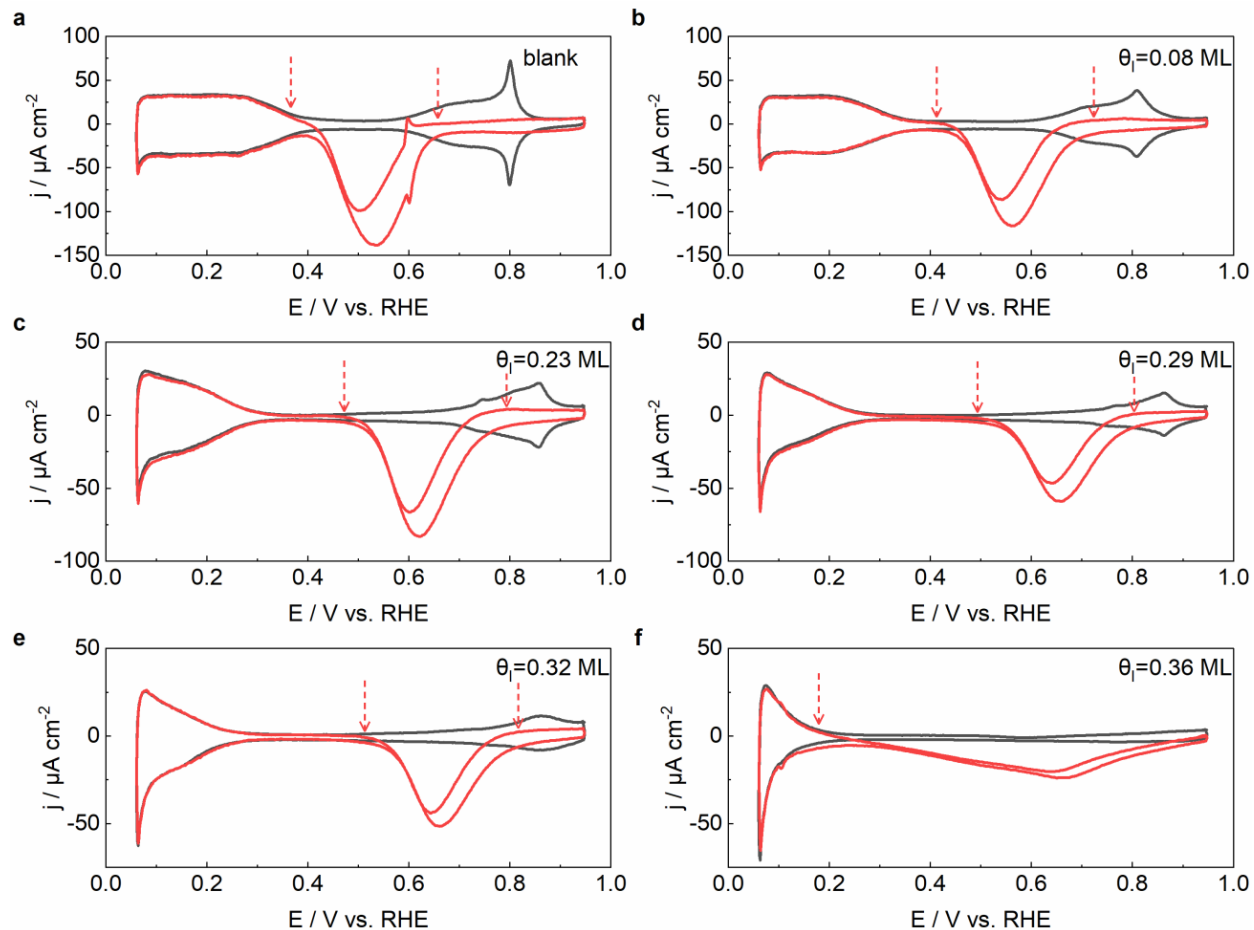

**Figure S13.** (**a-f**) Peroxodisulfate reduction (red) on P(111)@x ML I* electrodes in 0.1 M HClO$_4$ + 1 mM K$_2$S$_2$O$_8$. The scan rate is 50 mV/s. The blank voltammograms are also shown (black). The arrows indicate the potential of zero free charges.

The PZFC of clean Pt(111) and Pt(111)@I* with varying $\theta_I$ in HClO$_4$ solution, can be identified using S$_2$O$_8^{2-}$ (Peroxodisulfate, PDS) reduction as the probe reaction (Figure S13). Both the higher and a lower PZFC shift positively at $\theta_I$ < 0.32 ML, and shift negatively at $\theta_I$ > 0.32 ML (Figure S13). On clean Pt(111), the lower PZFC is 0.362 V, in excellent agreement with previously reported value at 0.370 V.[10] From $\theta_I$ = 0.0, i.e. clean Pt(111) to $\theta_I$ = 0.32 ML, the PZFC is shifted upward by ca. 150 mV. Considering adsorbed iodine atoms transfer a small of amount charge to Pt, this increase in PZFC must be due to a decrease in H$_2$O adsorption.[11, 12] When $\theta_I$ exceeds 0.32 ML, the PZFC begins to decrease with increasing coverage. This is probably due to the weak adsorption of water renders the electronic effect of I* on Pt dominating. Due to the relatively large size of the $S_2O_8^{2-}$ ion, the reduction current becomes very small when the iodine coverage is too high. Therefore, we only measured data within the range of 0 - 0.36 ML.

## 9. The ORR activity of Pt(hkl)@I* and polycrystalline Pt@I* electrodes and their kinetic current density.

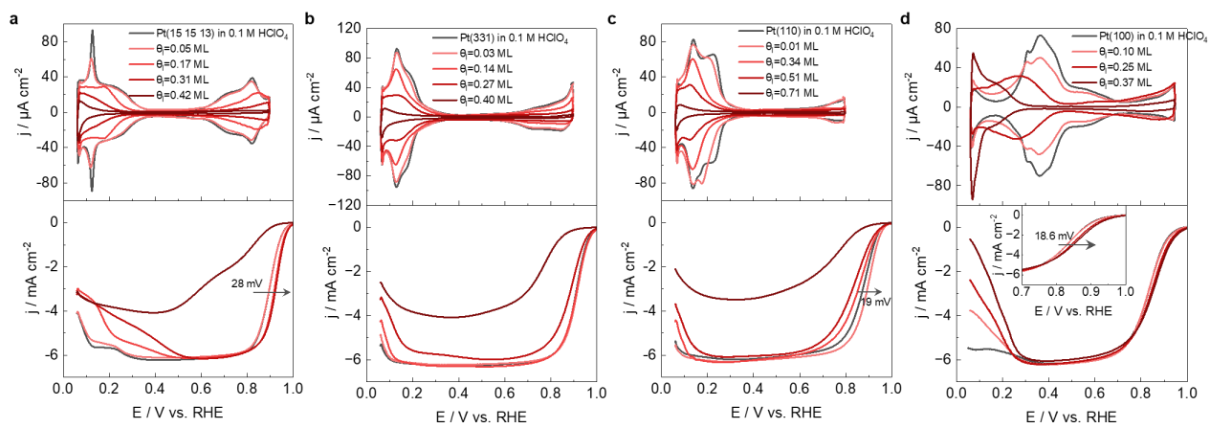

**Figure S14**. CVs and ORR polarization curves recorded in the positive scan in 0.1 M HClO$_4$ of (**a**) Pt(15 15 13)@ x ML I*, (**b**) Pt(331)@ x ML I*, (**c**) Pt(110)@ x ML I* and (**d**) Pt(100)@ x ML I*oelectrodes For comparison, the base CV and the j-E curve for ORR at bare Pt(hkl) are also included in (**a-d**). The scan rate is 50 mV/s. For the ORR experiments, the rotation rate is 1600 rpm.

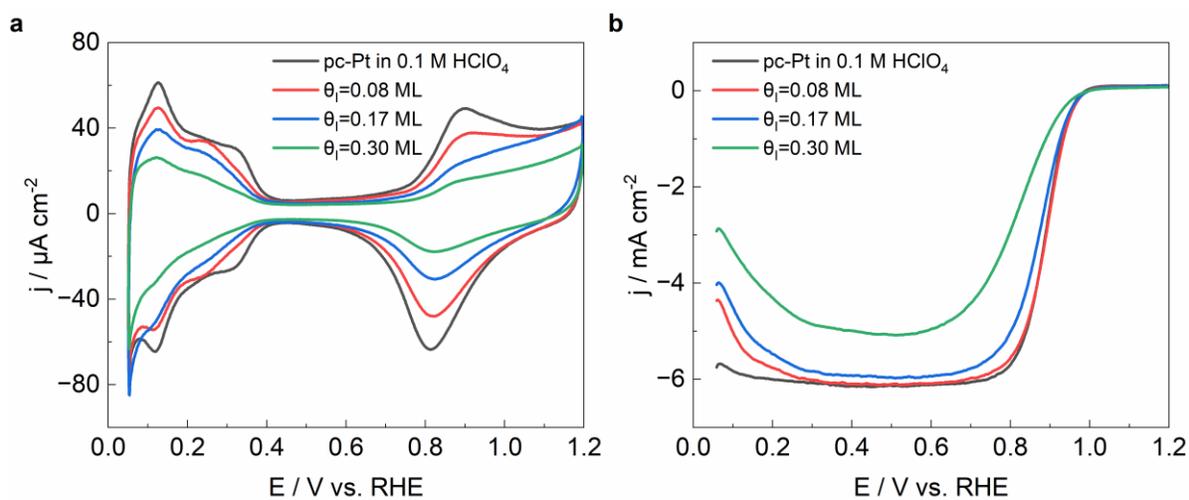

**Figure S15.** (**a**) CVs of polycrystalline Pt@I* electrodes in Ar-saturated 0.1 M HClO$_4$ with various coverages as indicated in the figure. (**b**) Positive-going j-E curves for ORR at Polycrystalline Pt@I* electrodes in O$_2$-saturated 0.1 M HClO$_4$. For comparison, the base CV and the j-E curve for ORR at bare polycrystalline Pt are also included in (**a**) and (**b**). Scan rate: 50 mV/s. For the ORR experiments, the rotation rate is 1600 rpm.

**10. The Bader charge of iodine adsorbed on the Pt surface were calculated using DFT, as well as the influence of iodine on the d-band center of Pt.**

Table S1. The average Bader charge of adsorbed I under different iodine coverage degrees.

| The coverage of I / % | The average Bader charge of each I |
|---|---|
| 6.3 | 0.0849 |
| 12.5 | 0.0783 |
| 18.8 | 0.0622 |
| 25.0 | 0.0422 |
| 31.3 | 0.0222 |
| 37.5 | 0.0269 |
| 43.8 | 0.0132 |

Table S2. The d-band center of Pt under different I coverage.

| The coverage of I / % | The d-band center of the first layer of Pt/ eV | The d-band center of all Pt/ eV |
|---|---|---|
| 0.0 | -1.9570 | -2.243 |
| 6.3 | -2.0030 | -2.234 |
| 12.5 | -2.1270 | -2.289 |
| 18.8 | -2.1720 | -2.278 |
| 25.0 | -2.1720 | -2.327 |
| 31.3 | -2.2000 | -2.294 |
| 37.5 | -2.3760 | -2.408 |
| 43.8 | -2.2990 | -2.334 |